\documentstyle[preprint,aps]{revtex}
\draft
\begin{document}
\title{ Wigner Crystals Phases in Bilayer Quantum Hall Systems}
\author{Subha Narasimhan and Tin-Lun Ho}
\address{Physics Department,  The Ohio State University, Columbus, Ohio
43210}
\maketitle

\vspace{.3in}

\centerline{\bf Abstract}

\vspace{0.2in}

Electrons in double-layer quantum well systems behave like pseudo-spin 1/2
particles where the up and down ``spin" represent localized
states in each of the layers.
The magnetically induced Wigner crystals in these systems are therefore
crystals of these pseudo-spin 1/2 particles.
We have calculated the phase diagram of the bilayer Wigner crystals using
a variational scheme which explores a
 $continuum$ of lattice and spin structure. Five stable
crystal phases are found.
For given tunneling strength and layer separation, one typically encounters
the following sequence of transitions as the filling factor is increased
from zero,
(the same sequence also occurs if
one increases the ``effective" layer separation starting
from zero, with tunneling strength and filling factor held fixed) :
\begin{tabbing}
\hspace{0.2in} \=
({\bf I}) (One-component) hexagonal structure \=
$\rightarrow$ \= ({\bf II}) Centered rectangular structure \\
$\rightarrow$  \> ({\bf III}) Centered square structure \>
$\rightarrow$  \> ({\bf IV}) Centered rhombic structure \\
$\rightarrow$  \> ({\bf V}) Staggered hexagonal structure. \>  \>
\end{tabbing}
Crystal {\bf I} is a ferromagnet in pseudo-spin space. All other
crystals ({\bf II} to {\bf V}) have mixed ferro- and antiferromagnetic
orders, which are generated by layer tunneling and interlayer repulsion
respectively. The relative strength of these two magnetic orders
vary continuously with external parameters,
(i.e. the ratio of layer separation to magnetic length,
the tunneling gap to
Coulomb interaction, etc).
The lattice structures  {\bf I}, {\bf III}, {\bf V}
are ``rigid" whereas {\bf II} and {\bf V} are ``soft", in the sense that
the latter two vary with external parameters and the former three do not.
Another important feature of the phase diagram is  the existence
a multicritical point and a critical end-point, which allows all crystals
(except {\bf V}) to transform into one another continuously.
While our findings are based on a variational calculation, one can
conclude on physical grounds that the
mixed ferromagnetic-antiferromagnetic  order as well as  the
pseudospin-lattice coupling should be general features of most
bilayer Wigner crystals.

\vspace{0.2in}


\pagebreak

\section{ Introduction}

Recent experiments\cite{ATT1} have shown that
bilayer quantum well systems in strong magnetic
fields behave like quantum Hall systems with  (pseudo-)spin 1/2,
where the up and down
pseudo-spins correspond to electron states residing in one or the
other of the layers. The novelty of this new quantum Hall system
is demonstrated by the quantum Hall effect at filling factors
$\nu=1/2$ and $\nu=1$, which corresponds to each layer being 1/4
filled and 1/2 filled respectively\cite{ATT1}.
Since even denominator quantum Hall fluids do not exist in
fully spin polarized single-layer systems, these states can only result
from the correlations between electrons in different layers.

One major difference between single-layer and bilayer systems is their
energy scales. In the single-layer case, if one restricts to the lowest
Landau level, the only remaining energy scale is the Coulomb interaction
between electrons $e^{2}\sqrt{n}$, where $n$ is the electron density.
($n$ is related to the filling factor $\nu$ as $n=\nu/(2\pi\ell^{2})$, where
$\ell=\left(c\hbar/eB\right)^{1/2}$ is the magnetic length.)
   For bilayer systems, there are three energy
scales: (a) the tunneling energy $\Delta$ between the layers, (precise
definition given later), (b) the inter-layer Coulomb interaction $e^{2}/D$,
where $D$ is the separation between the layers  and, (c) the
intra-layer Coulomb
interaction $e^{2}\sqrt{n}$, where $n=\nu/(2\pi\ell^{2})$
is the total electron density of $both$ layers.
Because of these energy scales, the system can fall into different physical
regimes  depending on their relative ratios. The important physical
regimes are:

\noindent (i) Two single-layer
regime, $e^{2}\sqrt{n} \gg e^{2}/D, \Delta$ ;

\noindent (ii) Two-component (or correlation) regime, $e^{2}\sqrt{n} \sim
e^{2}/D \gg \Delta$ ;

\noindent (iii) One-component regime, $e^{2}\sqrt{n} \sim e^{2}/D
< \Delta$.

Regime (i) corresponds to the case of large layer spacing. In this case, the
systems reduces to two weakly interacting single-layer systems.
Regime (iii) corresponds to sufficiently small layer spacing so that
the effects of tunneling becomes important. In this case, the electrons
lie in the ``symmetric" state of the quantum well. The system reduces again
to the single-layer case. Regime (ii) is very different. Despite  the
weakness of layer tunneling, electrons are
strongly coupled to each other through the Coulomb interaction.
The newly discovered $\nu=1/2$ and $\nu=1$ quantum Hall states are
found in this regime\cite{ATT1}. In particular, the $\nu=1/m$ states ($m$ odd)
are believed to behave like quantum ferromagnets\cite{QHF}.
[ By one and  two-component, we  have in mind a
pseudo-spin 1/2
representation of the electron  wavefunction.  We shall discuss this
representation shortly. ]

For single-layer systems, it is well known that the quantum Hall fluids
 eventually lose their stability at sufficiently low filling factors
to the Wigner crystals, which are essentially  products of electron
Gaussians arranged on a lattice, properly antisymmetrized to satisfy
the Pauli principle. The emergence of Wigner crystals at low filling is
inevitable, as electron exchange has  weakened so much, that the correlation
energy of the fluid falls below the Madelung energy of the Wigner solid.
In the last few years, many experiments (transport\cite{wc}\cite{wcbell},
threshold field\cite{thre}, magnetophonon\cite{magmode},
and luminlescence\cite{lum}) have indicated the existence of
crystalline characteristics at various filling factors,
some are as high as around $\nu=1/5$\cite{wc}\cite{wcbell}.
While there are questions as to
whether the insulating states around $\nu=1/5$
are Wigner crystals or other kinds of novel insulators\cite{ins},
the general view is that the insulating states at lower filling factors
will be Wigner crystals.

The  reasons for the appearance of Wigner crystal mentioned above
also apply to the bilayer case. However, because of the additional
internal degrees of freedom, bilayer systems have a
much greater variety of  Wigner crystal states. A limited version of this
rich variety can be seen even without calculations, which we present
schematically in figure 1.
When the layers are far apart, [i.e. regime (i)], the system
becomes two single layers, each one  has its own hexagonal Wigner
crystal.  To reduce interlayer repulsion, these two crystals will be
``staggered" as in the usual hexagonal close pack, (i.e.  the lattice points
of one lattice lies directly above the centers of the triangles of the other).
As the layer spacing $D$ is reduced, one enters regime (ii). In this regime,
 the staggered hexagonal structure in regime (i)  cannot survive  because
the large inter-layer repulsion will push the electrons to fill up the
``vacancies" in the staggered structure.
The simplest staggered crystalline structure without vacancies
is the staggered square lattice shown in figure 1,  where
the lattice sites of one lattice sit above the centers of the other.
As $D$ decreases further, one enters regime (iii). As mentioned before,
the bilayer system reduces to the single-layer case.
The crystal structure is therefore hexagonal.

The purpose of this paper is to discuss how these  crystals
transform into one another as the energy parameters and the filling factor
of the system are varied. We have performed a
variational calculation  which examines a large class
of lattice and pseudo-spin structures. The resulting phase diagram
is surprisingly rich.  It contains more crystalline phases
than the ones mentioned above.
It also contains a multicritical point and a critical end-point.
A key feature of these crystal phases is that their spin structure vary
over most parts of the phase diagram, even in those regions where the lattice
structure remains fixed.
In fact, as we shall see,
the simple-minded staggered square structure mentioned above
(which has fixed spin structure in the pseudo-spin language) is not
the optimal structure in general.
Another  common feature is that all bilayer crystals
acquire a net magnetization in the psuedo-spin space, except for the case
of $strictly$ zero tunneling. The existence of this
magnetization has important implications on the macroscopic properties of
 these electron solids, which we shall discuss elsewhere.

The rest of this paper is organized as follows. In Section II, we
describe our model and our variational wavefunction. The result of our
calculation, i.e. the Wigner crystal phase diagram, is presented in
Section III.  In Section IV, we describe in detail
our calculational scheme. Although our variational scheme is conceptually
straightforward, its actually implementation is not. The reason is that
in evaluating the energy of a crystal structure, one has to evaluate many
lattice sums which are very slowly converging. Moreover,  one has to
repeat the energy calculation thousands of times in the minimization scheme.
It is therefore necessary to develop tricks to speed up the evaluations of
the lattice sums to the point that makes the minimization process feasible.
When all the tricks are put together, the calculation is quite involved.
To make the presentation as clear as possible, we first describe the
computational strategy in detail in Section IV. The details of
evaluating the  relevant expressions are given in the Appendices.
At the end of Section IV, we also compare our
results with the recent work of some others on the same subject.
In Section V, we summarize our results, and  emphasize those
important features of bilayer Wigner crystals which we believe should be model
independent.

\bigskip

\section{The Wigner Crystals in Bilayer Quantum Hall Systems}

\noindent \underline{The model and the pseudo-spin representation} :

Consider the double-layer quantum well  shown in figure 2. We shall
adopt the coordinate system shown in figure (2a) and (2b).
The origin of coordinate
system is placed at the center of the well. The layers are parallel to
the $x$-$y$ plane.  The position of an electron will be denoted as
${\bf x}= ({\bf r}, z)$, where ${\bf r}=(x,y)$ denotes a two dimensional
vector. We shall consider magnetic fields {\bf H} normal the layers, i.e. along
${\bf \hat{z}}$.
The Hamiltonian of a system with $N$ electrons is
$H=H_{o}+V- B$, where $H_{o}$ is the single particle Hamiltonian, $V$
is the electron-electron interaction and $B$ is the (divergent)
classical contribution in $V$ (which we shall define later).
Explicitly, $H_{o}=\sum_{i=1}^{N}[ H_{||}({\bf r}_{i})+H_{\perp}(z_{i})]$,
where
$H_{||}$ describes the cyclotron motion in the
xy plane, and  $H_{\perp}$ describes the motion along $z$,
\begin{equation}
H^{o}_{||}({\bf r}) =  \frac{1}{2m^{\ast}}\left(\frac{\hbar}{i}
\frac{\partial}{\partial {\bf r}} -
\frac{e}{2c}{\bf H}\times{\bf r}\right)^{2}, \hspace{0.3in}
H_{\perp}(z) = -\frac{\hbar^{2}}{2m^{\ast}}\frac{{\rm d}^{2}}{dz^{2}} + U(z),
\end{equation} $m^{\ast}$ is the effective mass, and
$U(z)$ is a double well potential as shown in figure 2.
We shall consider the case where the ground state and first excited state
of $U(z)$ (denoted as $f_{+}$ and $f_{-}$ respectively) consists of
maxima at $D/2$ and $-D/2$. (See figure 2). The separation $D$ between
the maxima  will be referred to as
the ``separation" between the layers.

The  electron-electron interaction $V$ is given by
\begin{equation}
V = \sum_{i>j} \frac{e^{2}}{|{\bf x}_{i}-{\bf x}_{j}|}.
\label{vee} \end{equation}
Since the electrons are localized around $z=\pm D/2$, a system will total
electron density $n$ will contain a
``classical" contribution $B$, which is the Coulomb energy
of two infinitely thin layers located at $\pm D/2$, each with density $n/2$,
\begin{equation}
B =  \frac{e^{2}}{2}
\int {\rm d}^{3}{\bf y} {\rm d}^{3}{\bf y}'
\frac{n_{b}({\bf y})n_{b}({\bf y}')}{|{\bf y}-{\bf y}'|},
\end{equation}
\begin{equation}
n_{b}({\bf r}, z) = \frac{n}{2}\left[ \delta(z-D/2) + \delta(z+D/2)\right]
\end{equation}

The ground state $f_{+}$ and the first excited state $f_{-}$
 are symmetric and
antisymmetric about the center of the well. Their energies
will be denoted by $\epsilon_{+}$ and $\epsilon_{-}$.
For later discussions, we define
\begin{equation}
\Delta = \epsilon_{-} - \epsilon_{+},  \label{tun} \end{equation}
\noindent which we shall refer to as the ``tunneling gap".
The ground state of $H_{||}$ is the lowest Landau level with energy
$\frac{1}{2}\hbar\omega_{c}$, $\omega_{c}=eH/m^{\ast}c$, with wavefunctions
\begin{equation}
u_{m}({\bf r}) = \frac{([x+iy]/\ell)^{m}}{(2\pi 2^{m} m!\ell^{2})^{1/2}}
e^{-r^{2}/4\ell^{2}}
\label{lo}
\end{equation}
For large magnetic fields and a sufficiently narrow well,
it is sufficient to consider the lowest Landau level $\{ u_{m}(x,y)\}$
and the ``pseudo-spin 1/2" space spanned  by
$f_{+}$ and $f_{-}$. Although $f_{+}$ and $f_{-}$ are eigenstates of
$H_{\perp}$, sometimes it is more convenient to use the ``localized" basis
$f_{\uparrow}$ and $f_{\downarrow}$
\begin{equation}
f_{\uparrow} = \frac{1}{\sqrt{2}}\left( f_{+} + f_{-} \right) ,
\hspace{0.5in}
f_{\downarrow} = \frac{1}{\sqrt{2}}\left( f_{+} - f_{-} \right) ,
\label{fch} \end{equation}
\noindent which are states localized in the upper and lower layers
respectively, (see figure 2). If we define spinors
\begin{equation}
f(z) = \left(\begin{array}{c} f_{\uparrow}(z)  \\
f_{\downarrow}(z) \end{array} \right), \hspace{0.5in}
\zeta = \left(\begin{array}{c} \zeta_{\uparrow}\\ \zeta_{\downarrow}
\end{array} \right) ,  \label{fvec}  \end{equation}
\noindent a general linear combination of $f_{+}$ and $f_{-}$ can be
written as (using the fact that $f$ is real)
\begin{equation}
f^{\dagger}\cdot \zeta  \equiv \zeta_{\uparrow} f_{\uparrow} +
\zeta_{\downarrow} f_{\downarrow} , \label{fdot}
\end{equation}
\noindent which is completely specified by the spinor $\zeta$.

In defining the spinor $f$ in eq.(\ref{fvec}),
we have in fact implicitly chosen a coordinate system
$({\bf \hat{x}}_{1}, {\bf \hat{x}}_{2}, {\bf \hat{x}}_{3})$
in the pseudo-spin space so that $f_{\uparrow}$ and $f_{\downarrow}$ are
eigenstates of $\sigma_{3} = \vec{\sigma}\cdot{\bf \hat{x}}_{3}$, and
$f_{+}$ and $f_{-}$    are
eigenstates of $\sigma_{1} = \vec{\sigma}\cdot{\bf \hat{x}}_{1}$
respectively,
\begin{equation}
f_{\uparrow} \rightarrow \left(\begin{array}{c} 1\\0\end{array}\right) ,
\hspace{0.4in}
f_{\downarrow} \rightarrow \left(\begin{array}{c} 0\\1 \end{array}\right) ,
\hspace{0.4in}
f_{+} \rightarrow \frac{1}{\sqrt{2}}
\left(\begin{array}{c} 1\\1 \end{array}\right) ,
\hspace{0.4in}
f_{-} \rightarrow \frac{1}{\sqrt{2}}
\left(\begin{array}{c} 1\\-1 \end{array}\right).
\label{fcor} \end{equation}
\noindent There are no relations between
$({\bf \hat{x}}_{1}, {\bf \hat{x}}_{2}, {\bf \hat{x}}_{3})$
and the real space directions ${\bf \hat{x}}$,
${\bf \hat{y}}$, and ${\bf \hat{r}_{\perp}}$.
In general, a  spinor can be written as
\begin{equation}
\zeta = \left(\begin{array}{c} {\rm cos}\frac{\theta}{2} e^{-i\phi/2} \\
{\rm sin}\frac{\theta}{2} e^{i\phi/2} \end{array} \right) e^{-i\chi/2} ,
\label{zgen} \end{equation}
\noindent  where $\theta$ and $\phi$ are the polar angles of its
spin vector ${\bf S}$.
\begin{equation}
{\bf S} = \zeta^{\dagger}{\vec \sigma}\zeta =
{\rm cos}\theta {\bf \hat{x}}_{3} +
{\rm sin}\theta \left(  {\rm cos}\phi\  {\bf \hat{x}}_{1}
+ {\rm sin}\phi\ {\bf \hat{x}}_{2} \right).
\label{spin} \end{equation}
\noindent
The spin vectors of $f_{+}, f_{-}, f_{\uparrow}$, and $f_{\downarrow}$ are
${\bf \hat{x}}_{1}, -{\bf \hat{x}}_{1}, {\bf \hat{x}}_{3}$ and
${\bf \hat{x}}_{3}$ respectively. (See fig. 2). In this representation,
the energy of $N$ non-interacting electrons in
the lowest Landau level and in the pseudo-spin space is
\begin{equation}
H_{o} = - {1\over 2}\Delta S_{1} ,  \hspace{0.4in}
S_{1} = {\bf S}\cdot {\bf \hat{x}}_{1} =
\sum_{i=1}^{N} S_{1,i} .
\label{hspin} \end{equation}
\noindent where we have ignored the constant $N(\hbar\omega_{c}/2)$.

\vspace{0.2in}

\noindent \underline{Bilayer Wigner Crystal Variational States } :

In the single-layer case, Wigner crystals can be constructed using
coherent states\cite{MZ}.
A coherent state at ${\bf R}$ is
defined as
\begin{equation}
\phi_{{\bf R}}({\bf r}) = \frac{1}{\sqrt{2\pi \ell^{2}}}
{\rm exp}\left( -\frac{r^{2}}{4\ell^{2}} +
\frac{(x+iy)(R_{x}-iR_{y})}{2\ell^{2}}
-\frac{R^{2}}{4\ell^{2}} \right) \equiv <{\bf r}|{\bf R} >
\label{coh}
\end{equation}
A simple variational state
for the Wigner crystal  can be obtained
by antisymmetrizing a product of coherent states distributed on a
lattice $\{ {\bf R}_{i}\}$. To simplify the notation, we write
\begin{equation}
\phi_{{\bf R}_{i}}({\bf r}_{j}) \equiv \phi_{i}(j) \label{sim} \end{equation}
\noindent The Wigner crystal wavefunction is then
\begin{eqnarray}
&|\Phi_{wc}> &=  {\cal A}\left[ \prod^{N}_{i=1}|i>\right] =
\sum_{P} (-1)^{P}\left[ \prod_{i=1}^{N}|Pi>\right].
\label{cwc}  \\
&\Phi_{wc}([{\bf r}]) &=  {\cal A}
        \left[ \prod^{N}_{i=1}\phi_{i}(i)\right] =
\sum_{P} (-1)^{P}
\left[ \prod_{i=1}^{N}\phi_{Pi}(i)\right]
\end{eqnarray}
\noindent where $[{\bf r}]\equiv ({\bf r}_{1}, {\bf r}_{2}, ...,
{\bf r}_{N})$, ${\cal A}$ is an
antisymmetrizer for the electron coordinates $\{ {\bf r}_{j} \}$,
and $P$ denotes a
permutation of $N$ objects with signature $(-1)^{P}$.

Bilayer coherent states are simply single-layer ones augmented by
a pseudo-spin structure,
\begin{equation}
\phi_{_{{\bf R},\zeta}}({\bf x}) = \left(f^{\dagger}(z)
\cdot\zeta\right) \phi_{\bf R}({\bf r})  \equiv <{\bf x}|{\bf R}, \zeta>,
\label{prod} \end{equation}
\noindent recalling that ${\bf x} = ({\bf r}, z), {\bf r}= (x,y)$.
The analog of the Wigner crystal of the single-layer case can be obtained
by antisymmetrizing the product of a set of $N$ coherent states,
distributed on a regular array of $N$ points $\{ {\bf R}_{i}, i=1, ..N \}$
in the $x$-$y$ plane.
Each point ${\bf R}_{i}$ is associated with a spinor
$\zeta({\bf R}_{i})$ describing the wavefunction of this coherent state along
$r_{\perp}$. As in the single-layer case, we write \begin{equation}
\phi_{i}({\bf j}) \equiv \phi_{{\bf R}_{i},\zeta({\bf R}_{i})}({\bf x}_{j}).
\label{msim} \end{equation}   The index $i$ now stands for both
the 2D lattice site ${\bf R}_{i}$ and the spinor  $\zeta({\bf R}_{i})$,
where the bold
face ${\bf j}$ denotes the three dimensional coordinate ${\bf x}_{j}$ of
the $j$-th electron. In this notation,
Wigner crystal wavefunctions are still of the form eq.(\ref{cwc}). More
explicitly, they are
\begin{equation}
\Phi_{wc}([{\bf x}]) = {\cal A}
\left[ \prod^{N}_{i=1} \phi_{i}({\bf i})\right] =
\sum_{P} (-1)^{P} \left[ \prod_{i=1}^{N}\phi_{Pi}({\bf i}) \right]
\label{wcwf}
\end{equation}
\noindent Since these crystals are completely specified by
the two dimensional
array $\{ {\bf R}_{i} \}$ and the spinors $\{ \zeta({\bf R}_{i}) \}$, it is
to useful to represent them as ``two-dimensional" crystals of ``spin 1/2"
particles. In this representation, the
hexagonal structure in regime (iii) (i.e. figure 1c) corresponds to the
ferromagnetic state shown in figure 3a, with  magnetization  along
${\bf \hat{x}}_{1}$, representing the symmetric state along $r_{\perp}$.
The centered square (fig. 1b) and staggered hexagonal (fig. 1c) structures
mentioned in Section I correspond to the antiferromagnetic
structures shown in figure 3b and 3c. The spins along
 $+{\bf \hat{x}}_{3}$ or $-{\bf \hat{x}}_{3}$ in these figures correspond
to electrons in the upper or lower layer.

Ideally, one would like to do a variational calculation taking  the set
$\{ {\bf R}_{i} \}$ and the spinors $\{ \zeta({\bf R}_{i})\}$ as variables.
Such a parameter space is too large to be practical. To anticipate the
effects of tunneling on the lattice and the spin structure we have
discussed, we restrict ourselves to the following
configurations:
We shall consider systems with an even number of electrons, ($N$ even).
The array $\{ {\bf R}_{i}\}$  consists of two lattices
A and B, which are
identical except shifted relative to one another by a vector ${\bf c}$.
Spinors on the same lattice are identical, but need not be the same
on different lattices.  In other words, if ${\bf a}_{1}$ and
${\bf a}_{2}$ are basis vectors of A, then
\begin{eqnarray}
&{\bf R} &=  n_{1}{\bf a}_{1} + n_{2}{\bf a}_{2}, \hspace{0.2in}
 {\rm and} \hspace{0.2in} \zeta({\bf R})=\zeta_{A} , \hspace{0.2in} {\rm if}
\hspace{0.2in} {\bf R} \in A,
\nonumber \\
&{\bf R} &=  m_{1}{\bf a}_{1} + m_{2}{\bf a}_{2} + {\bf c}, \hspace{0.2in}
 {\rm and} \hspace{0.2in} \zeta({\bf R})=\zeta_{B} , \hspace{0.2in} {\rm if}
\hspace{0.2in} {\bf R} \in B.
\label{defa}
\end{eqnarray}
\noindent where $\{ n_{i}, m_{i}\}$ are integers.

Our variational calculation is performed at fixed electron density $n$ and
fixed magnetic field. The variational parameters are the
spinors $\zeta_{A}$ and $\zeta_{B}$, the displacement vector ${\bf c}$,
and the basis vectors ${\bf a}_{1}, {\bf a}_{2}$. The latter are
subject to the constant area constraint
$|{\bf a}_{1}\times{\bf a}_{2}|=2/n$.
The energy function to be minimized is
\begin{equation}
E = {\rm Min}_{_{\Psi}} \left( \frac{<\Psi| \left( V - B \right)|\Psi>}
{<\Psi| \Psi>} -  \Delta S_{1} \right)
\label{fun}
\end{equation}
\noindent In the single-layer case, since there is only one energy scale,
$e^{2}/\ell$,  the energy per particle is of the form
\begin{equation}
\frac{E}{N} = \frac{e^{2}}{\ell}{\cal E}(\nu). \end{equation}
where ${\cal E}$ is a dimensionless function of $\nu$.
In the bilayer case, because of the three energy scales
 $e^{2}\sqrt{n}$, $e^{2}/D$, $\Delta$, the energy per particle
is of the form
\begin{equation}
E = \frac{e^{2}}{\ell} {\cal E}(D/\ell, \Delta/(e^{2}/\ell), \nu)
\end{equation}
The labor of our  calculation is to minimize the energy eq.(\ref{fun})
within the variational space
$\{ \Gamma = (\zeta_{A}, \zeta_{B}, {\bf c}, {\bf a}_{1}, {\bf a}_{2}) \}$.
The phase diagram of the system is given by the
optimal configuration $\Gamma^{o}$ as a function of the experimental
parameters $\left\{D/\ell,  \Delta/(e^{2}/\ell), \nu\right\}$.

In the next section, we shall present the Wigner crystal phase diagram
according to our calculation. The details of our calculation
will be  given in Section IV.

\bigskip

\section{Phase diagram}

Our variational calculation reveals an unexpected Wigner crystal phase diagram.
Because of its richness,  we shall display it in two different ways.
We  first show the phase diagram in the plane of
$\Delta/(e^{2}/a)$ and $D/a$ for various
filling factors $\nu$, (see figures 4a, 4b, and 4c), where
\begin{equation}
a^{2} = 2/n =4\pi/\nu \end{equation}
is the area of the unit cell of lattice A.
Since the layer spacing $D$ and the tunneling gap $\Delta$
are fixed in actual experiments, variations in $\Delta/(e^{2}/a)$ and
$D/a$ correspond to varying the electron density $n$  (say, by varying the
gate voltages). Figures 4a, 4b, and 4c are for $\nu= 1/3, 1/5$, and 0
respectively.
Of course, at $\nu=1/3$ and $1/5$, the system is a quantum Hall fluid. The
reason that we still choose to display the Wigner crystal phase diagram at
these fillings is because it is  essentially unchanged at nearby fillings,
where the system is no longer a quantum Hall fluid. In fact, when plotted
in terms of the variables $\Delta/(e^{2}/a)$ and $D/a$, the phase boundaries
show only relatively small shifts over a large range of filling factors.
On the other hand, if the phase diagram is displayed
in terms of the variable $\overline{\Delta}\equiv \Delta/(e^{2}/\ell)$ and
$\nu$ for different values of $D/\ell$,  (figures 5a to 5d), the movements of
the phase boundaries become  much more pronounced. Figures 5a to 5d will be
useful for experiments where electron densities are fixed so that
variations in magnetic field causes variations in $\overline{\Delta}$ and
$\nu$.

We have found altogether five stable Wigner crystal states.
We shall label them together with their region of stability as
{\bf I, II, III, IV, V}.  Roughly speaking,
the two single-layer regime (i.e. regime (i)) mentioned in Section I is
contained in region {\bf V}. The correlation regime (ii) is contained in region
{\bf II, III, IV}. The one-component regime (iii) is contained in region
{\bf I}. The optimal spin configuration turns out to be
\begin{equation}
{\bf S}_{A(B)} =  {\rm sin}\theta {\bf \hat{x}}_{1}
+ (-) {\rm cos}\theta {\bf \hat{x}}_{3} \label{spingen} \end{equation}
The spin structure of crystal {\bf I} is ferromagnetic,
(i.e. $\theta=90^{o}, {\bf S}_{A}={\bf S}_{B}={\bf \hat{x}}_{1}$).
All other crystals have mixed ferromagnetic and antiferromagnetic order,
with  $0\leq \theta < 90^{o}$. These system has a uniform magnetization
$2{\rm sin}\theta{\bf \hat{x}}_{1}$ and a staggered magnetization
$2{\rm cos}\theta{\bf \hat{x}}_{3}$. The pure antiferromagnetic case
$\theta=0$ only occurs at zero tunneling.
 The five classes of stable Wigner crystals are :

\noindent \underline{Region {\bf I} :  One-component ferromagnet
hexagonal crystals} :
The lattices A and B of these crystals are staggered
in such a way so that their union is a hexagonal lattice,
[${\bf a}_{1}\cdot{\bf a}_{2}=0,\ a_{2}/a_{1}=\sqrt{3},\
 a_{1}^{2}\sqrt{3}/2=1/n,\ {\bf c}=({\bf a}_{1}+{\bf a}_{2})/2$].
In the entire region {\bf I}, ${\bf S}_{A}={\bf S}_{B}={\bf \hat{x}}_{1}$,
i.e. all electrons are in the symmetric state in the
$r_{\perp}$ direction. This structure is ``rigid".
Both its lattice and spin structures are fixed in the entire region {\bf I}.

\noindent \underline{Region {\bf II} : Mixed ferromagnetic and
antiferromagnetic centered  rectangular crystals} :
Both sublattices  A and B are rectangular lattices,
with B sitting at the center of the unit cell of A:
$[{\bf a}_{1}\cdot{\bf a}_{2} = 0,\ a_{1}a_{2} = 2/n,\ {\bf c} =
({\bf a}_{1} + {\bf a}_{2})/2 ]$. Unlike the ``rigid" hexagonal structure
in region {\bf I}, these structures are ``soft" in the sense that
both the spin angle $\theta$ and the lattice parameter
$a_2/a_1$  vary continuously throughout the entire region {\bf II}.
(See figure 6a).
These crystals are separated
from the hexagonal ones  in region {\bf I} by a  first order line, which
changes into a second order line at a multicritical point.  It is also
separated from crystal {\bf III} mentioned below by a second order line, which
intersects the {\bf I}-{\bf II}
 first order line at a critical end-point.

\noindent \underline{Region {\bf III} : Mixed ferromagnetic and
antiferromagnetic centered square crystals }:
Both sublattices
A and B are square lattices. They stagger in the same way as those in
region {\bf II}. The lattice structure of these crystals are rigid, with
$[{\bf a}_{1}\cdot{\bf a}_{2}=0, a_{1}=a_{2}=\sqrt{2/n},
{\bf c}=({\bf a}_{1} + {\bf a}_{2})/2]$. However, the spin structure is soft.
The spin angle $\theta$ changes continuously within this region.
These crystals are separated from crystal {\bf I} by a first order line, and
crystal {\bf II} by a second order line.

\noindent \underline{Region {\bf IV} :  Mixed ferromagnetic and
antiferromagnetic centered  rhombic crystals} :
Both A and B are rhombic lattices. They stagger in the same way as crystal
{\bf I} and {\bf II}.  $[ a_1=a_2, \
a_1^2 \sin\alpha =2/n,
{\bf c}=({\bf a}_{1} + {\bf a}_{2})/2]$.
Like crystal {\bf II}, this structure is soft. The spin angle
$\theta$, and the lattice angle $\alpha$ between
${\bf a}_{1}$ and ${\bf a}_{2}$
vary throughout the entire region {\bf IV}. (See figure 6b).
These crystals are separated from crystal {\bf III} by a second order line,
and crystal {\bf V} mentioned below by a first order line.

\noindent \underline{Region {\bf V}: Mixed ferromagnetic and
antiferromagnetic staggered hexagonal crystals} :
Both A and B are hexagonal lattices.
A is sitting above the center of triangle of B. The lattice structure
is rigid, with $[a_{1}= a_{2},\
\alpha=60^{o},\ a_{1}^{2}\sqrt{3}/2 = 2/n,
{\bf c}=({\bf a}_{1} + {\bf a}_{2})/3]$. However, its spin structure is soft.
The angle $\theta$ varies over the entire region V.

Figures 7 and 8 show the variations of the spin angle along the horizontal
and vertical line in figure 5a.  There is a large change in spin angle
(about $20^{o}$) across the first order line separating the
one-component hexagonal and centered square structure.
However, the change across the centered rhombic and
staggered hexagonal first order line is small,
(about $2^{o}$).
As we shall explain in the next section, the spin angle
for all cases is given by :
\begin{equation}
{\rm sin}\theta=\cases{ 1 & if $\gamma\leq \overline{\Delta}$\cr
        \overline{\Delta} / \gamma
                & if $\gamma > \overline{\Delta}$\cr}
\label{angle}
\end{equation}
where $\overline{\Delta}=\Delta/(e^{2}/\ell)$, and $\gamma$ is a function
of the lattice structure $({\bf a}_{1}, {\bf a}_{2}, {\bf c})$,
$\nu$, and $D/\ell$,.
For crystals ${\bf I}, {\bf III},
{\bf V}$ where the lattice structure is rigid, (see descriptions
{\bf I}, {\bf III}, {\bf V}  above), $\gamma$ reduces to $\gamma^{{\bf I}},
\gamma^{{\bf III}}, \gamma^{{\bf V}}$ which are functions of
$\nu$ and $D/\ell$ only.
We have plotted these three curves in figures 9a to 9d.
Readers who are interested
in the spin angle $\theta$ in regions {\bf I}, {\bf III}, {\bf V} for given
$D/\ell$ can extract their  values from eq.(\ref{angle}) using these
curves.
The determination of $\theta$ in regions ${\bf II}$ and ${\bf IV}$ is less
straightforward, for the lattice structures are soft in these regions.
One has to first determine the lattice structure by numerical minimization,
and then evaluate $\gamma$ following the prescriptions in Section IV and
Appendices C and D.

\bigskip

\section{Variational Calculation }

In this section, we shall describe in detail our variational calculation.
Our goal is to minimize the energy per particle at fixed  densities
with respect to the variational Wigner crystal wavefunctions of
eq.(\ref{wcwf}).
\begin{equation}
{\cal E} = < H > /N = \left[ <H_{o}> + <V> - B \right]/N \equiv
 {\cal E}_{o} + {\cal V} - {\cal B}
\label{etotal}
\end{equation}
\noindent where $<\hat{O}> = <\Psi_{wc}|\hat{O}|\Psi_{wc}>
/<\Psi_{wc}|\Psi_{wc}>$.
To simplify the notation, we shall from now on
measure lengths in units of magnetic length $\ell$, and
energy in units of $e^{2}/\ell$. In these units, the tunneling energy is
\begin{equation}
{\cal E}_{o} =
-\frac{1}{4}\overline{\Delta}(S_{1}^{A}+S_{1}^{B}), \qquad
\overline{\Delta}\equiv\Delta/(e^{2}/\ell)
\label{hzero}
\end{equation}
Separating the background interaction $<B>$ into contributions due to
each layer, we have
\begin{equation}
{\cal B} = \frac{n}{4}\int \frac{{\rm d}^{2}y}{y}
+ \frac{n}{4}\int \frac{{\rm d}^{2}y}{\sqrt{y^{2} + D^{2}}  }
= - \left[ \frac{2\pi}{Ga^2}\Bigg|_{G\rightarrow 0}
- {\pi D\over a^2} \right]
\label{bsim} \end{equation}
where $a^2=2/n=4\pi/\nu$ is the unit cell area.
Although bilayer Wigner crystals are more complex than the single-layer
ones, they are  identical in form when represented as in eq.(\ref{cwc})
and eq.(\ref{wcwf}).
The evaluation of $<V>$ can therefore  proceed identically as in Maki
and Zotos\cite{MZ}. The result is an expansion in a set of $n$-body
potentials,
\begin{equation}
<V> = \sum_{i>j}V(ij) + \sum_{i>j>k}V(ijk) + \cdots
\label{mz}
\end{equation}
where
\begin{equation}
V(ij) = \frac{(<ij|-<ji|)V(|ij> -|ji>)}{2(1 - |<i|j>|^{2})}
\label{vij}
\end{equation}
Recalling that $i$ stands for ${\bf R}$ and $\zeta({\bf R})$, we have
\begin{eqnarray}
&&V(ij) = \frac{1}{2} \left(\frac{1}{1-|S(ij)|^{2}}\right)
\int {\rm d}^{3}{\bf x}
\int {\rm d}^{3}{\bf x}'\  \frac{1}{|{\bf x}-{\bf x}'|}
|\phi_{i}({\bf x})\phi_{j}({\bf x}')
-\phi_{j}({\bf x})\phi_{i}({\bf x}')|^{2}
\nonumber \\
&&\qquad\qquad \left|S_{ij}\right|^{2} =
\left|\int {\rm d}^{3}{\bf x}\
        \phi_{i}^{\ast}({\bf x})\phi_{j}({\bf x})
\right|^{2}
\label{vij1}
\end{eqnarray}
The integrals in eq.(\ref{vij1}) include both direct and exchange terms,
which are proportional to $|\phi_{i}({\bf x})\phi_{j}({\bf x}')|^{2}$
 and $\phi_{i}^{\ast}({\bf x})\phi_{j}({\bf x})
\phi_{j}^{\ast}({\bf x}')\phi_{i}({\bf x}')$ respectively. As in the
single-layer case, the exchange terms as well as the $n$-body term for
$n\leq 3$ are smaller than the direct terms by
a factor of ${\rm exp}[-R^{2}/2\ell^{2}]$, where $R=|{\bf R}_{i}-
{\bf R}_{j}|$ is the distance between two coherent states. This factor is
less than $10^{-3}$ for $\nu < 1/2$. Thus, if we focus on Wigner crystals at
filling factors less than 1/2, the exchange terms in eq.(\ref{vij1})
and the $n$-body terms in eq.(\ref{mz}) for $n\leq 3$
can be neglected.  $V(ij)$ then takes the Hartree form.
\begin{equation}
V(ij) = \int {\rm d}^{3}{\bf x}\
\int {\rm d}^{3}{\bf x}' \frac{1}{|{\bf x}-{\bf x}'|}
|\phi_{i}({\bf x})\phi_{j}({\bf x}')|^{2}, \hspace{0.2in} {\rm for}
\hspace{0.2in} \nu < 1/2
\label{hartree}
\end{equation}
Next, using
\begin{equation}
\frac{1}{|{\bf x} - {\bf x}'|} = \int \frac{{\rm d}^{2}q}{(2\pi)^{2}}
\frac{2\pi}{q} e^{i{\bf q}\cdot({\bf r}-{\bf r}')}
e^{-q|r_{\perp}-r_{\perp}'|}.
\label{fou}
\end{equation}
eq.(\ref{hartree}) can be written as
\begin{eqnarray}
&V(ij) &= \int_{0}^{\infty} {\rm d}q\ e^{-q^{2}} J_{o}(qR) W(q)
\label{simvij} \\
&W(q) &= \int {\rm d}u {\rm d}u'\ e^{-q|u-u'|}\
|f^{\dagger}(u)\cdot\zeta({\bf R}_{i})|^{2}\
|f^{\dagger}(u')\cdot\zeta({\bf R}_{j})|^{2} .
\label{wdef}
\end{eqnarray}
Recall that $f_{\uparrow}(r_{\perp})$ and $f_{\downarrow}(r_{\perp})$ are
localized around $r_{\perp}=D/2$ and $-D/2$. Typically, these
functions decay away from the layers within a decay length $1/\kappa$,
[i.e. $f_{\uparrow(\downarrow)}(r_{\perp})\sim {\rm exp}(-\kappa
|r_{\perp}-(+)D/2|)$].
In Appendix A, we show that as long as the products
 $\kappa D$ and $\kappa \ell$ are moderately larger than 1,
(referred the ``moderate" condition),
the spinor products in eq.(\ref{wdef}) can be replaced by
\begin{equation}  |f^{\dagger}(u)\cdot\zeta|^{2} =
|\zeta_{\uparrow}|^{2}\delta(u-D/2)
+ |\zeta_{\downarrow}|^{2}\delta(u+D/2). \label{appr} \end{equation}
These ``moderate" conditions are certainly feasible in experiments.
Within the approximation eq.(\ref{appr}), eq.(\ref{simvij})
reduces to the  simple form
\begin{eqnarray}
&V(ij)  &= U_{+}({\bf R}) + U_{-}({\bf R})S_{3}(i)S_{3}(j)
\equiv V_{S_{3}(i), S_{3}(j)}({\bf R})
\label{ss} \\
&U_{\pm}({\bf R}) &= {1\over 2}
        \int^{\infty}_{0} {\rm d}q\ e^{-q^{2}}\
(1 \pm e^{-qD})\ J_{o}(qR)
\label{upm}
\end{eqnarray}
where ${\bf R}={\bf R}_{i}-{\bf R}_{j}$, $S_{3}(i)\equiv
S_{3}({\bf R}_{i})$.

Note that the ineffectiveness of the exchange in eq.(\ref{vij1}) and
eq.(\ref{wdef}) does not mean that the problem is classical.
Quantum mechanical effects are manifested through (a) the smearing
of the (classical) $\delta$-function density to a Gaussian in the lowest
Landau level, (thereby affecting the interaction of two electrons at
distances of the order of $\ell$), and (b) layer tunneling, (i.e.
the tunneling gap $\Delta$ in eq.(\ref{hzero})).
[Note that the insignificance of layer exchange in eq.(\ref{wdef}) under
the ``moderate" condition does not mean that the tunneling gap in
eq.(\ref{hzero}) can also be ignored. The reason is that $\Delta$ is of the
order of $|\epsilon |e^{-\kappa D}$, $|\epsilon
|=|\epsilon_{+}+\epsilon_{-}|/2$.
Even though $e^{-\kappa D}$ is small when $\kappa D$ is moderately larger
than 1, $\Delta$ can still be comparable with with other energies in the
system for sufficiently attractive quantum well, which makes
$\epsilon$ sufficiently large.]

To evaluate ${\cal V}$, we separate the contributions from different
lattices A and B. Denoting $S_{3}^{A}$ and $S_{3}^{B}$ as $S$ and $S'$
respectively, we have
\begin{eqnarray}
&{\cal V}  &= \frac{1}{4}
        \sum_{{\bf R}\neq 0}
        \left[ V_{S, S}({\bf R}) + V_{S', S'}({\bf R})\right]
        + \frac{1}{2}\sum_{{\bf R}}
        V_{S, S'}({\bf R}+{\bf c})
\label{curryv} \\
&&\equiv  \eta -  \frac{\gamma}{4} \left(\frac{S^A_3 - S^B_3}{2}\right)^2
        +  \frac{\lambda}{4}\left(\frac{S^A_3 + S^B_3}{2}\right)^2
\label{disspin}  \\
&\eta &= {1\over2}\left[\sum_{{\bf R}\neq 0} U_{+}({\bf R})
+ \sum_{{\bf R}} U_{+}({\bf R} + {\bf c}) \right]
\label{defeta} \\
&\gamma &= -2\left[\sum_{{\bf R}\neq 0} U_{-}({\bf R})
- \sum_{{\bf R}} U_{-}({\bf R} + {\bf c})\right]
\label{defgamma} \\
&\lambda &= 2\left[\sum_{{\bf R}\neq 0} U_{-}({\bf R})
+ \sum_{{\bf R}} U_{-}({\bf R} + {\bf c})\right] .
\end{eqnarray}
All the spin dependence in the energy are contained in eq.(\ref{disspin})
and eq.(\ref{hzero}), which contains $(S^{A}_{3}, S^{B}_{3})$ and
$(S^{A}_{1}+S^{B}_{1})$ respectively. Since both ${\bf S}^{A}$  and
${\bf S}^{B}$ are vectors of fixed length,  it is easy to see that
the energy is minimized when  $S_{2}^{A}=S_{2}^{B}=0$.
Direct plotting shows that $U_{-}(R)$ is a positive,
monotonic decreasing function of $R$.
This means $\lambda$ is strictly
positive.  $\gamma$  may be positive or negative
depending on whether
${\bf c}$ is smaller or larger than the shortest vector in A.
In either case, it is straightforward to show that the optimum spin
structure is
\begin{equation}
S^{A}_{3} = - S^{B}_{3}= {\rm cos}\theta,\qquad
S_{1}^{A}=S_{2}^{B}={\rm sin}\theta
\end{equation}
Thus the total correlation energy per electron (eq.(\ref{etotal})  is
\begin{equation}
{\cal E} = \xi  +
{1\over 4}\gamma {\rm sin}^{2}\theta  -{1\over 2}
        \overline{\Delta} {\rm sin}\theta,
\hspace{0.3in} \xi = \eta - {1\over 4}\gamma - {\cal B}.
\label{final} \end{equation}
The optimum spin angle is therefore given by
\begin{equation}
{\rm sin}\theta = \cases{ 1 & if $\gamma \leq \overline{\Delta}$\cr
          \overline{\Delta}/\gamma  & if $\gamma >
\overline{\Delta}$ \cr}
\label{ang} \end{equation}
and the corresponding energies are
\begin{equation}
{\cal E} = \cases{ \xi + {1\over 4}\gamma -{1\over2}\overline{\Delta} &
                if $\gamma \leq \overline{\Delta}$ \cr
                \xi -\overline{\Delta}^{2}/(4\gamma) &
                if $\gamma > \overline{\Delta}$ \cr}
\end{equation}
Thus determination of the correlation energy and the optimum spin
structure for a given lattice reduces to the evaluation $\xi$
and $\gamma$.
To evaluate these quantities, it is useful to introduce
the following sums. Let us define the functions
\begin{eqnarray}
&F_{1}(R) &\equiv \int^{\infty}_{0} {\rm d}q\  e^{-q^{2}}J_{o}(qR),
\nonumber \\
&F_{2}(R) &\equiv \int^{\infty}_{0} {\rm d}q\ e^{-qD}e^{-q^{2}}J_{o}(qR)
\end{eqnarray}
and the lattice sums
\begin{equation}
Q_{i} = \sum_{{\bf R}\neq 0} F_{i}(|{\bf R}|),
\qquad\qquad
\overline{Q}_{i} = \sum_{{\bf R}} F_{i}(|{\bf R}+{\bf c}|),
\qquad\qquad  i=1,2.
\end{equation}
The functions $\eta$
and $\gamma$, (hence $\xi$ and $\gamma$),  can now be expressed as
\begin{eqnarray}
&\eta - {\cal B} &= {1\over 4}\left[Q_{1} + \overline{Q}_{1} +
Q_{2} + \overline{Q}_{2}\right] - {\cal B}
\label{etab} \\
&-\gamma &= (Q_{1} - \overline{Q}_{1})
        - (Q_{2}-\overline{Q}_{2})
\label{another}
\end{eqnarray}
The evaluation of $\eta - {\cal B}$ and $\gamma$ (or $\xi$ and $\gamma$)
reduces to the evaluation of the four
sums $\{Q_{i}, \overline{Q}_{i}, i=1,2\}$.
The reason that we consider the particular combination $\eta-{\cal B}$ is
because (as we shall see) each $Q$ term has a divergent (classical)
contribution
so that their total contribution is ${\cal B}$. As a
result, the combination $\eta-{\cal B}$ is finite. Likewise, the
differences $(Q_{i}-\overline{Q}_{i}), i=1,2$ are all finite because their
divergent contributions cancel each other.

A straightforward forward evaluation of
these sums proves highly impractical as they converge very slowly. On the
other hand, an essentially exact evaluation is possible if one notes the
following : (i)
The asymptotic forms of  $F_{i}(R), i=1,2$
consist only of powers in $1/R$ or $1/\sqrt{R^2+D^2}$
(see Appendix B),
(ii) all sums of the form
$\sum_{{\bf R}\neq 0}R^{-(2n+1)}$ and
$\sum_{R} 1/(R^2+D^2)^{n+1/2}$ (where $n$ is a positive integer)
can be either evaluated analytically or transformed into a rapidly
convergent series convenient for  numerical calculation. (See Appendix C).
We therefore adopt the following strategy :

\noindent (a)  For any desired accuracy, (which we choose to be $10^{-9}$),
we choose a cutoff $\Lambda$ and a set of asymptotic functions
$\{F^{as}_{i}(R), i=1,2\}$ which are obtained by retaining a few terms
of the asymptotic expansions of $F_{i}(R)$ such that
$\{F_i^{as}(R)\}$ reduce to $\{F_{i}(R)\}$ within
the desired accuracy for all $R>\Lambda$. (See Appendix B).
Next, we rewrite the lattice sums $Q_{i}$ and $\overline{Q}_{i}$ as
\begin{equation}
Q_{i} =  Q^{sr}_{i} + Q^{as}_{i}, \hspace{0.5in}
\overline{Q}_{i} =  \overline{Q}^{sr}_{i} + \overline{Q}^{as}_{i},
\label{qsum} \end{equation}
where
\begin{equation}
Q^{as}_{i} = \sum_{{\bf R}\neq 0} F^{as}_{i}(R),
\hspace{0.4in}
\overline{Q}^{as}_{i} = \sum_{{\bf R}} F^{as}_{i}(|{\bf R}+{\bf c}|),
\label{Qasdef} \end{equation}
and
\begin{eqnarray}
&Q^{sr}_{i} &= \sum_{{\bf R}\neq 0}\left[F_{i}(R) -
        F^{as}_{i}(R)\right]\Theta(\Lambda-R)
\nonumber \\
&\overline{Q}^{sr}_{i} &= \sum_{{\bf R}} \left[
        F_{i}(|{\bf R}+{\bf c}|) - F^{as}_{i}(|{\bf R}+{\bf c}|)
        \right] \Theta(\Lambda-R)
\end{eqnarray}
The superscipt $sr$ means ``short range", and $\Theta(x) = 1$ or $0$
if $x>$ or $<0$. It is clear that all the short range contributions
$\{Q^{sr}_{i}, \overline Q_i^{sr}\}$ are finite.

\noindent (b) The sums $\{Q^{as}_{i}, \overline Q_i^{as}, i=1,2\}$
are evaluated analytically by generalizing the method of
Bonshall and Maradudin\cite{BM}, or transforming  the
sum into a  very rapidly convergent series.
These are done in detail in Appendix C and D. As we shall see in Appendix C,
all four asymptotic sums  contain a divergent piece
typical of Madelung sums. Separating out this divergent
piece, they can all be written as
\begin{eqnarray}
&Q^{as}_{1} &= \frac{2\pi}{Ga^2}\Big|_{G\rightarrow 0} + T_{1}, \qquad
Q^{as}_{2} = \frac{2\pi}{Ga^2}\Big|_{G\rightarrow 0}
        -{2\pi D\over a^2} + T_{2}
\nonumber \\
&\overline{Q}^{as}_{1} &= \frac{2\pi}{Ga^2}\Big|_{G\rightarrow 0} +
       \overline{T}_{1},  \qquad
\overline{Q}^{as}_{2} = \frac{2\pi}{Ga^2}\Big|_{G\rightarrow 0}
        - {2\pi D\over a^2}+ \overline{T}_{2}
\label{deft2b}
\end{eqnarray}
where the $\{ {T}_{i}, \overline{T}_{i}\}$ are the finite parts  of
$\{Q^{as}_{i}, \overline{Q}^{as}_{i} \}$.
Using eq.(\ref{etab}) and (\ref{another}), $\xi$ and $\gamma$ in
eq.(\ref{final}) are now expressed in terms of the finite quantities
$\{ T_{i}, \overline{T}_{i}, Q^{sr}_{i}, \overline{Q}^{sr}_{i}\}$:
\begin{eqnarray}
&\xi &= (\eta - {\cal B}) -\frac{1}{4}\gamma \nonumber \\
&\eta - {\cal B} &= \sum_{i=1,2} {1\over 4}
\left[ Q^{sr}_{i} + \overline{Q}^{sr}_{i} + T_{i} +
        \overline{T}_{i}\right] \nonumber \\
&-\gamma &= \left[ (Q^{sr}_{1}-\overline{Q}^{sr}_{1}) -
        (Q^{sr}_{2}-\overline{Q}^{sr}_{2})
        + (T_{1}-\overline{T}_{1}) -
        (T_{2}-\overline{T}_{2}) \right]
\end{eqnarray}
The explicit expressions of $T_{i}$ and $\overline{T}_{i}$ are given in
Appendix C.

\noindent (c) The short range sums $Q^{sr}_{1}$ and $\bar{Q}^{sr}_{1}$
can be evaluated directly. This is because  the asymptotic
function $F^{as}_{1}(R)$, is a simple sum of inverse powers of $R$,  and
$F_{1}(R)$ is proportional to the zeroth order
order Bessel function with imaginary argument, which is available
in most math libraries.
The evaluation of $\{ Q^{sr}_{2}, \overline{Q}^{sr}_{2}\}$  is less
straightforward as the integral $F_{2}(R)$ is  not a tabulated
special function. Although this function can be evaluated to high
accuracy by numerical integration, incorporating this integration
in the minimization process is too time consuming to be practical.
However, this bottleneck can be eliminated by the following trick.
We first evaluate $F_{2}(R)$ by numerical integration
on a fine mesh within
the short range region $R<\Lambda$. The values of $F_{2}(R)$ at
any point not on the mesh can be obtained (to a desired accuracy)
from the nearby mesh points by using the ``cubic spline" interpolation
\cite{num}. This allows us to replace all necessary numerical
integrations in our minimization process by the  spline interpolates,
which is enormously faster.

With the short range contributions given by (c), and the asymptotic
contributions given in Appendix C and D, our evaluation of the functions
$\xi$ and $\gamma$ is complete.

\vspace{0.5in}

{\bf The accuracy of the present calculation :} The most serious
approximation in our calculation is that the exchange terms are ignored
in eq.(\ref{vij1}). As mentioned at the beginning of this section,
the exchange terms are smaller than the direct terms by a factor
of $e^{-R^{2}/2\ell^{2}}$, which is $10^{-3}$ around $\nu=1/2$ and
drops dramatically at lowering fillings, ($10^{-5}$ around $\nu=1/3$ and
$10^{-7}$ around $\nu=1/5$). Once the exchange terms are dropped, the
rest of our variational calculation is essentially excact. The
numerical scheme we have mentioned can easily acheive an accuracy of
one part in $10^{9}$, (and can be improved systematically).
To demonstrate further the accuracy of our calculation, we show in Figure
10(a) and 10(b) the energies of vairous crystal states at $\nu=1/3$.
The exchange energy is invisible on the scale of both figures.
One can clearly see on figure 10 that the energy differences between
different crystals range from $10^{-1}$ (far from the phase boundary)
to $10^{-2}$ (close to the phase boundary), except for those between
crystal {\bf I} and {\bf II},
and between {\bf II} and {\bf III}. The latter is shown on an expanded scale in
figure 10(b), from which one can see that the energy difference of these
crystals is typically of the order to $10^{-3}$. Thus, the energy
differences between all crystals are much greater than the exchanged
energy $10^{-5}$. Furthermore, the exchange energy appears as a
{\em systematic correction} to all crystals. In comparing the energies between
crystals of similar crystal structures, they are to a large extent
cancelled out so that the actual exchange
contributions to the energy $difference$
are at least an order of magnitute smaller than
$10^{-5}$.  The only place where our calucaltion may begin to produce a
few percent error bar is close to $\nu=1/2$. At lower fillings, all
numerical evidence indicate that our evaluations of crystal energies
are accurate to
the order of $e^{-\pi/\nu}$, (which is the ratio between exchange and
direct terms), and that the energy difference has an even higher
accuracy $(10^{-1} e^{-\pi/\nu})$ because of the aforementioned systematic
correction.

\bigskip

\section{Concluding Remarks :}

We have seen that Wigner crystals in bilayer quantum Hall systems come in
different ``magnetic" and structural varieties. Although our conclusions are
based on a variational calculation, there are two characteristics of these
crystals which are direct consequences of the
competition between layer tunneling and interlayer Coulomb repulsion, and
should therefore exist in real systems.
They are the coexistence of antiferromagnetic and ferromagnetic order, and the
coupling between lattice and spin structure. The antiferromagnetism in
pseudo-spin space is to minimize the
interlayer Coulomb repulsion, whereas the ferromagnetism is to
 take advantage of the tunneling energy. The coupling between
lattice and spin configurations simply reflects the competition
of these two energies.
 In many ways, Wigner crystals in bilayer
systems are like $^{3}$He solids, which also have different phases with
different lattice and magnetic structures.
The magnetic structures in $^{3}$He solids are known to give rise to
many remarkable nuclear magnetic resonances.
It will not be surprising if bilayer Wigner crystals also have many interesting
``magnetic" properties.

\bigskip

\section{Achnowledgement}

Part of this work was done during a three month's visit of T.L. Ho to
the National High Magnetic Field Laboratory at Tallahassee in Winter 94,
and during the 1994 Physics Summer School at the Hong Kong University of
Science and Technology.
He would like to
thank Nick Bonesteel and Bob Schrieffer for discussions, and specially
C. Jayaprakash for discussions on
multicritical points and critical end-points.
This work is
supported in part by NHMFL through NSF Grant DMR 9016241.

\vspace{1.0in}

\noindent \underline{{\bf Appendix A}.  Conditions for eq.(\ref{appr}) }:

The integrand in eq.(\ref{wdef}) contains terms like
$f_{\uparrow}(u)^{2}f_{\downarrow}(u')^{2}$,
$f_{\uparrow}(u)f_{\downarrow}(u)f_{\downarrow}(u')^{2}$,
$f_{\uparrow}(u)^{2}f_{\uparrow}(u')^{2}$, etc. We shall discuss the
contributions of  the first two terms, from which the contributions of the
rest can be similarly inferred. To simplify these expressions, we shift the
origins of $u$ and $u'$ so that the maxima of
$f_{\uparrow}$ and $f_{\downarrow}$ are at the origin,
\begin{equation}
f_{\uparrow}(u) = g(u-D/2) = f_{\downarrow}(-u) .
\label{shift} \end{equation}
The contribution of
$f_{\uparrow}(u)^{2}f_{\downarrow}(u')^{2}$ and
$f_{\uparrow}(u)f_{\downarrow}(u)f_{\downarrow}(u')^{2}$ to eq.(\ref{wdef})
can then be written as
\begin{equation}
I_{1}(q) = \int {\rm d}s \int {\rm d}s' e^{-q|D+s-s'|}g(s)^{2}g(s')^{2}
\label{ione} \end{equation}
\begin{equation}
I_{2}(q) = \int {\rm d}s \int {\rm d}s' e^{-q|D+s-s'|}g(s)g(s+D)g(s')^{2}
\label{itwo} \end{equation}
Since $g(s) \sim e^{-\kappa |s|}$, the exponential factor in eq.(\ref{ione})
and (\ref{itwo}) can be replaced by $e^{-q(D+s-s')}$ if $\kappa D$ is
moderately larger than 1. On the other hand, the Gaussian factor in
eq.(\ref{simvij}) limits $q$ to  the range $q\leq 1/\ell$. Hence, if
$\kappa\ell$ is moderately larger than one so that  $e^{-qs}$
decays much slower that $e^{-\kappa s}$, the functions $g(s)^{2}$ and
$g(s')^{2}$ in eq.(\ref{ione}) will act like $\delta$-functions.
The same  approximations applies to $I_{2}$. However, because of
the product, $I_{2}$ is smaller than $I_{1}$ by a factor
of $e^{-\kappa D}$. It can therefore be ignored if $\kappa D$ is
moderately larger than 1. Applying the analysis to other products
of $f$'s, we have eq.(\ref{appr}).

\vspace{1.0in}

\noindent \underline{{\bf Appendix B} : Asymptotic expansion of $F_{1}(R)$ and
$F_{2}(R)$} :

\vspace{0.2in}

Note that
\begin{equation}
F_{1}(R) = \int^{\infty}_{0}{\rm d}q e^{-q^{2}}J_{o}(qR) =
\frac{\sqrt{\pi}}{2} I^{e}_{o}(R^{2}/8). \end{equation}
where $I^{e}_{o}(x) = e^{-x}I_{o}(x)$, and $I_{o}$ is the Bessel function
with imaginary argument. For $R\gg1$, $F_{1}$ has an asymptotic expansion
\begin{equation}
F_{1}(R) \rightarrow F_{1}^{\infty}({\bf R})
= \sum_{n=0}^{\infty} \frac{1}{n!}\nabla^{2n}\frac{1}{R}
= \sum_{n=0}^{\infty} \frac{[(2n-1)!!]^{2}}{n!} \frac{1}{R^{2n+1}}
\end{equation}
For $R>\Lambda=35$, (i.e. $\Lambda=35\ell$ in the original unit),
only three terms in the asymptotic series
\begin{equation}
F^{as}_{1}(R) = \left[ 1 + \nabla^{2} + \frac{1}{2}\nabla^{4} \right]
\frac{1}{R}
\end{equation}
is enough to reproduce $F_{1}(R)$ to the accuracy of $10^{-9}$.

In the case of $F_{2}(R)$, we note that it has the asymptotic expansion
\begin{equation}
F_{2}(R) \to F_2^\infty(R) = \sum_{n=0}^{\infty}
\left[ \frac{(-1)^{n}}{n!} \frac{{\rm d}^{2n}}{{\rm d}D^{2n}}
\right] \frac{1}{\sqrt{R^{2}+D^{2}}} ,
\end{equation}
\begin{equation}
= \sum^{\infty}_{n=0} (-1)^{n} \frac{(2n)!}{n!}
\frac{1}{ (R^{2}+D^{2})^{n+1/2} }P_{2n}\left[ \frac{D}{\sqrt{R^{2}+D^{2}}}
\right] , \end{equation}
$P_{n}(x)$  are  the Legendre polynomials.
As in the case of $F_{1}$, for $R>\Lambda=35$, only three terms
in the above series
\begin{equation}
F^{as}_{2}(R) = \left[ 1 - \frac{{\rm d}^{2}}{{\rm d}D^{2}}
+ \frac{1}{2}\frac{{\rm d}^{4}}{{\rm d}D^{2}} \right]
\frac{1}{\sqrt{R^{2}+D^{2}}}
\end{equation}
is enough to reproduce $F_{1}(R)$ to the accuracy of $10^{-9}$.
$P_{n}(x)$  are  the Legendre polynomials.
The functions $F^{as}_{1}(R)$ and $F^{as}_{2}(R)$ are the asymptotic
functions we have used in our numerical  calculation.

\pagebreak

\noindent \underline{{\bf Appendix C} : Explicit determination of the
asymptotic lattice sums $\{ Q_i^{as}, \overline{Q}_i^{as},i=1,2\}$}:

\noindent Consider the following  generating functions :
\begin{eqnarray}
&h({\bf c}) &= \sum_{{\bf R}} \frac{1}{|{\bf R} + {\bf c}|}
\nonumber \\
&g({\bf c}, D) &= \sum_{{\bf R}}
\frac{1}{(|{\bf R} + {\bf c}|^{2}+D^2 )^{1/2}}
\end{eqnarray}
where ${\bf R}\in A$.
Using the asymptotic functions $\{F_1^{as}(R), F_2^{as}(R)\}$
defined in Appendix B, it is easy to see that
\begin{eqnarray}
&Q^{as}_{1} &=
{\rm Lim}_{{\bf b}\rightarrow {\bf 0}}
\left[ 1 + \nabla_{{\bf b}}^{2} + \frac{1}{2}
        \nabla_{{\bf b}}^{4} \right]
\left( h({\bf b}) - \frac{1}{b} \right) \nonumber \\
&\overline Q^{as}_{1} &=
        \left[ 1 + \nabla_{{\bf c}}^{2} + \frac{1}{2}
        \nabla_{{\bf c}}^{4} \right] h({\bf c}) \nonumber \\
&Q^{as}_{2} &=
        \left[ 1 - \frac{{\rm d}^{2}}{{\rm d}D^{2}}
        + \frac{1}{2}\frac{{\rm d}^{4}}{{\rm d}D^{4}} \right]
        \left( g({\bf 0},D) - {1\over D} \right) \nonumber \\
&\overline Q^{as}_{2} &=
        \left[ 1 - \frac{{\rm d}^{2}}{{\rm d}D^{2}}
        + \frac{1}{2}\frac{{\rm d}^{4}}{{\rm d}D^{4}} \right]
        g({\bf c},D)
\label{defqas}
\end{eqnarray}
We shall first give the expressions of $h$ and $g$. Their derivations
 are given in Appendix D.  These expressions are
\begin{eqnarray}
&h({\bf c}) &= \frac{2\pi}{Ga^2}\Bigg|_{G\rightarrow 0} +{1\over a}\left[
        -2 + L(c/a)\right] + {1\over a}\sum_{{\bf R}\neq 0}
\left[ L(|{\bf R}+{\bf c}|/a)+L(R/a){\rm cos}(2\pi|{\bf R}\times{\bf c}|/a^2)
\right] \label{defh} \\
&g({\bf c}, D) &=
\frac{2\pi}{Ga^2}\Bigg|_{G\rightarrow 0} - {2\pi D\over a^2}
+ {1\over a}\left[
\sum_{{\bf R}\neq 0} \frac{e^{-2\pi R D/a^2}}{(R/a)}
{\rm cos}(2\pi|{\bf R}\times{\bf c}|/a^2)  \right] \label{defg}
\end{eqnarray}
where $a^{2}$ is the unit cell area of lattice A,
($a^{2}= 2/n = 4\pi\ell^{2}/\nu$),  and
\begin{equation}
L(x) = \frac{1}{x}\left( 1- \phi(\sqrt{\pi} x) \right),  \hspace{0.3in}
\phi(y) = \frac{2}{\sqrt{\pi}}\int^{y}_{0} {\rm d}u\  e^{-u^{2}}
\label{defl}
\end{equation}
The function $\phi$ is the error function.

It may seem that only eq.(\ref{defg}) is necessary because
$h({\bf c})$ is a limiting case of $g({\bf c},D)$.
The latter is true but impractical. The reason is that the
number of terms needed to include
in the sum in eq.(\ref{defg}) to achieve a specified accuracy grows as
$1/D$ as $D\rightarrow 0$.  The series in eq.(\ref{defg}) is therefore
useless in the small $D$ limit. We are, however, lucky for two reasons.
First of all,
the typical value of $D$ in real experiments is of order  unity.
The series expansion eq.(\ref{defg}) is  highly convergent for
these $D$ values. As for $h({\bf c})$ [which corresponds to the special
case $g({\bf c}, D=0$)], an analytic expression  [eq.(\ref{defh})]
can be obtained by a straightforward
generalization of the method of Bonshall-Maradudin (BM)\cite{BM}.
(See appendix D).
This method produces a super convergent series  for
$h$, (a cutoff of just 4 lattice constants in the sum is
usually sufficient to produce an
accuracy of 10$^{-12}$).

Next, we note that the gradients in eq.(\ref{defqas})
can be conveniently evaluated using the following identities:
\begin{eqnarray}
&\nabla^{2}L(r) \equiv L_2(r) = \left[ L(r) + p_{1}(\pi r^{2})
        e^{-\pi r^{2}}\right]/r^2 ,
\hspace{0.3in} &p_{1}(x) = 4x + 2  \nonumber \\
&\nabla^{4}L(r) \equiv L_4(r) =  \left[ 9L(r) + p_{2}(\pi r^{2})
        e^{-\pi r^{2}} \right]/r^4 ,
\hspace{0.3in} &p_{2}(x) = 16x^{3}-8x^{2}+12x+18
\end{eqnarray}

\noindent
With these identities, using the definitions in eq.(\ref{deft2b}),
and eq.(\ref{defqas}) it is straightforward to work out the finite parts
$\{T_{i}, \overline{T}_{i}, i=1,2\}$ of the asymptotic sums
$\{Q^{as}_{i}, \overline{Q}^{as}_{i}, i=1,2\}$.
For $Q^{as}_{1}$, we have
\begin{eqnarray}
&& {\rm Lim}_{ {\bf b}\to {\bf 0}}  \left( h({\bf b})-1/b\right)
        \equiv \frac{2\pi}{Ga^{2}}\Bigg|_{G\to0} + T_{1a} \nonumber \\
&&{\rm Lim}_{{\bf b}\to {\bf 0}} \nabla^2_{{\bf b}}
   \left(h({\bf b})-1/b\right) \equiv T_{1b} \nonumber \\
&&{1\over 2}{\rm Lim}_{{\bf b}\to {\bf 0}} \nabla^4_{{\bf b}}
   \left(h({\bf b})-1/b\right) \equiv T_{1c} \nonumber \\
&&T_1 \equiv T_{1a} + T_{1b} + T_{1c}
\end{eqnarray}
where
\begin{eqnarray}
&T_{1a} &= {1\over a} \left[ -4  +
        2\sum_{{\bf R\ne0}} L(R/a)\right] \nonumber \\
&T_{1b} &= {1\over a^3} \left[ {8\pi\over 3}  +
        \sum_{{\bf R\ne0}} \left(
        L_2(R/a) - 4\pi^2 (R/a)^2 L(R/a) \right)
        \right]\nonumber \\
&T_{1c} &= {1\over 2a^5} \left[ -{64\pi^2\over 5}  +
        \sum_{{\bf R\ne0}} \left( L_4(R/a) +
        16\pi^4  (R/a)^4 L(R/a) \right)
        \right]
\end{eqnarray}

\bigskip
\noindent Similarly, for $\overline{Q}^{as}_{1}$, we have
\begin{eqnarray}
&&h({\bf c}) \equiv \frac{2\pi}{Ga^{2}}\Bigg|_{G\to0}+
        \overline T_{1a},
\qquad  \nabla^2_{{\bf c}}  h({\bf c})
        \equiv \overline{T}_{1b},
\qquad  {1\over 2} \nabla^4_{{\bf c}}  h({\bf c})
        \equiv \overline{T}_{1c}
\nonumber \\
&&\overline{T}_{1} \equiv
        \overline{T}_{1a}+\overline{T}_{1b}+\overline{T}_{1c}
\end{eqnarray}
where
\begin{eqnarray}
&\overline T_{1a} &= {1\over a} \left[ L(c/a)  +
        \sum_{{\bf R\ne0}} \left( L(|{\bf R+c}|/a) +
        L(R/a)\cos (2\pi |{\bf R\times c}|/a^2)
        \right) \right] \nonumber \\
&\overline T_{1b} &= {1\over a^3} \left[ L_2(c/a)  +
        \sum_{{\bf R\ne0}} \left( L_2(|{\bf R+c}|/a) -
        4\pi^2  (R/a)^2 L(R/a) \cos (2\pi |{\bf R\times c}|/a^2)
        \right) \right]\nonumber \\
&\overline T_{1c} &= {1\over 2a^5} \left[ L_4(c/a)  +
        \sum_{{\bf R\ne0}} \left( L_4(|{\bf R+c}|/a) +
        16\pi^4  (R/a)^4 L(R/a)\cos (2\pi |{\bf R\times c}|/a^2)
        \right) \right]
\end{eqnarray}

\bigskip
\noindent Next, for $Q^{as}_{2}$, we have
\begin{eqnarray}
&& \left[ g({\bf 0},D)-1/D\right]
        \equiv \left(\frac{2\pi}{Ga^{2}}
        \Bigg|_{G\rightarrow 0}-\frac{2\pi D}{a^{2}}\right)
        + T_{2a},\nonumber \\
&&  -{{\rm d}^2\over {\rm d}D^2} (g({\bf 0}, D) - 1/D)
        \equiv T_{2b},\nonumber \\
&&   {1\over 2}{{\rm d}^4\over {\rm d}D^4} (g({\bf 0}, D) -1/D)
        \equiv T_{2c} \nonumber \\
&&T_2 \equiv  T_{2a} + T_{2b} + T_{2c}
\end{eqnarray}
where
\begin{eqnarray}
&T_{2a} &=  {1\over a} \left[
        {-1\over (D/a)} + \sum_{{\bf R}\leq 0}
        \frac{e^{-2\pi RD/a^{2}}}{R/a} \right] \nonumber \\
&T_{2b} &= -{1\over a^3}\left[ - {2!\over (D/a)^3} +
          \sum_{{\bf R\ne0}} 4\pi^2 (R/a) e^{-2\pi D R/a^2}
        \right] \nonumber \\
&T_{2c} &= {1\over 2a^5} \left[
        - {4!\over (D/a)^5} +
          \sum_{{\bf R\ne0}} 16\pi^4 (R/a)^3 e^{-2\pi D R/a^2}
         \right]
\end{eqnarray}

\bigskip
\noindent Finally, for $\overline{Q}^{as}_{2}$, we have
\begin{eqnarray}
&&g({\bf c}, D) \equiv
        \left(\frac{2\pi}{Ga^2}\Bigg|_{G\rightarrow 0}
        -\frac{2\pi D}{a^2}\right) + \overline T_{2a},
        \nonumber \\
&& -{{\rm d}^2\over {\rm d}D^2} g({\bf c}, D)
        \equiv \overline T_{2b},
        \nonumber \\
&& {1\over 2} { {\rm d}^4 \over {\rm d}  D^4} g({\bf c}, D)
        \equiv \overline T_{2c} \nonumber \\
&&\overline T_2=\overline T_{2a} +
        \overline T_{2b}+\overline T_{2c}
\end{eqnarray}
where
\begin{eqnarray}
&\overline T_{2a} &= {1\over a}
        \sum_{{\bf R}\neq 0} \frac{e^{-2\pi R D/a^2}}{(R/a)}
        {\rm cos}(2\pi|{\bf R}\times{\bf c}|/a^2),
        \nonumber \\
&\overline T_{2b} &= -{1\over a^3}
        \sum_{{\bf R\ne0}} 4\pi^2 (R/a) e^{-2\pi D R/a^2}
        \cos( 2\pi |{\bf R\times c}|/a^2),
        \nonumber \\
&\overline T_{2c} &= {1\over 2a^5}
        \sum_{{\bf R\ne0}} 16\pi^4 (R/a)^3 e^{-2\pi D R/a^2}
        \cos( 2\pi |{\bf R\times c}|/a^2).
\end{eqnarray}

\bigskip
\noindent A final note ---
to obtain an overall accuracy in the correlation energy to eight
significant digits, suitable cutoffs ($\Lambda_{h}$ and $\Lambda_{g}$)
for the lattice sums in
eq.(\ref{defh}) and (\ref{defg}) are
$\Lambda_{h} \approx 5a$ and $\Lambda_{g}\approx 20a^2/\pi D\ $
(so that $e^{-2\pi D\Lambda_h/a^{2}} \sim 10^{-11}$
in eq.(\ref{defg})).

\vspace{1.0in}

\noindent \underline{{\bf Appendix D}:  Derivation of $h({\bf c})$ and
$g({\bf c}, D)$.}

\noindent
The series for $h({\bf c})$ in eq.(\ref{defh})
can be derived by a straightforward
generalization of the BM method\cite{BM}.  We first write
\begin{equation}
h({\bf c}) = \sum_{{\bf R}} \frac{2}{\sqrt{\pi}}
\left(\int^{w}_{0}  + \int^{\infty}_{w} \right) {\rm d}\beta\
e^{-\beta^{2}|{\bf R}+{\bf c}|^{2}}  \equiv (i) + (ii)
\label{onetwo}
\end{equation}
where $w=\sqrt{\pi}/a$.  After rescaling
the integration variable, the second term $(ii)$ can be written as
\begin{equation}
(ii) = \frac{1}{a}\sum_{{\bf R}} L(|{\bf R} + {\bf c}|/a),
\label{res3}
\end{equation}
where $L(R)$ is defined in eq.(\ref{defl}).
To evaluate the first term $(i)$, we first convert the
real space lattice sum into a sum in the reciprocal space,
(using the standard relation $\sum_{{\bf R}} \omega({\bf R})
=\sum_{{\bf G}}\tilde{\omega}({\bf G})/a^2$, where $\tilde{\omega}$
is the Fourier transform of $\omega$),
\begin{equation}
(i) = \sum_{{\bf G}} \frac{2}{\sqrt{\pi}} \int^{w}_{0} {\rm d}\beta\
\frac{\pi}{a^2\beta^{2}}
e^{-G^{2}/4\beta^{2} + i {\bf G}\cdot{\bf c}}
\end{equation}
Changing $\beta \rightarrow w/\beta$, we have
\begin{equation}
(i)= \frac{2}{a}\sum_{{\bf G}} \int^{\infty}_{1} {\rm d}\beta \
e^{-(G\beta/2w)^{2} + i{\bf G}\cdot{\bf c}}
\equiv \sum_{{\bf G}} I_{{\bf G}}
\end{equation}
The ${\bf G}=0$ term is  connected to the $1/R$ divergence of
the sum and has to be treated separately:
\begin{equation}
I_{{\bf G}=0} = \frac{2}{a}{\rm Lim}_{{\bf G}\rightarrow 0}
\left[ \int^{\infty}_{0} - \int^{1}_{0}\right] {\rm d}\beta\
e^{-(G\beta/2w)^{2} + i{\bf G}\cdot{\bf c}}
= \frac{2\pi}{Ga^2}\Bigg|_{G\rightarrow 0} - \frac{2}{a}
\label{res2}
\end{equation}
For the ${\bf G}\neq 0$ terms, we note that reciprocal lattice vectors
${\bf G}$ and real space lattice vectors ${\bf R}$ are related as
${\bf G} = 2\pi {\bf \hat{z}}\times {\bf R}/a^2$. We can then write
\begin{eqnarray}
&\sum_{{\bf G}\neq 0} I_{{\bf G}} &= \frac{1}{a} \sum_{{\bf R\ne0}}
        2\int^{\infty}_{1} {\rm d}\beta\
        e^{-\pi R^2\beta^{2}/a^2 + i 2\pi |{\bf R}\times {\bf c}|/a^2}
\nonumber \\
&&= \frac{1}{a}\sum_{{\bf R\ne0}} L(R/a)\
        \cos (2\pi|{\bf R}\times {\bf c}|/a^2)
\label{res1}
\end{eqnarray}
Substituting eqs.(\ref{res3}), (\ref{res2}), and (\ref{res1}) into
eq.(\ref{onetwo})
we have eq.(\ref{defh}).

\bigskip
The derivation of eq.(\ref{defg}) for $g({\bf c},D)$ is as follows.
We write $g({\bf c},D)$ as
\begin{equation}
g({\bf c}, D) = \sum_{{\bf R}} \int {{\rm d}^2q\over (2\pi)^2}
        {2\pi\over q}\ e^{i{\bf q\cdot (R+c)}}\ e^{-qD}
\end{equation}
Using the relation
\begin{equation}
\sum_{{\bf R}} e^{i{\bf q\cdot R}} = \sum_{{\bf G}}
        (2\pi)^2 \delta({\bf q}-{\bf G})/a^2
\end{equation}
we find
\begin{equation}
g({\bf c}, D) =   \sum_{{\bf G}} {2\pi \over Ga^2}
        e^{i{\bf G\cdot c}-GD}
\end{equation}
which is eq.(\ref{defg}) if we treat the ${\bf G=0}$ term separately, and
and write ${\bf G}= 2\pi{\bf \hat z\times R}/a^2$.

\pagebreak

\bigskip

\noindent {\bf Figure Captions}

\vspace{0.2in}

\noindent Figure 1 : (1a) Staggered hexagonal structure in regime (i);
(1b) Centered square structure in regime (ii);
(1c) Single component hexagonal structure in regime (iii).
The electrons are in the symmetric state of the double quantum well, which
are represented schematically by a ``peanut" shape.

\vspace{0.2in}

\noindent Figure 2 : A simple model of the double layer quantum well and
the pseudo-spin representation for the lowest doublet.
The coordinate system and the quantum well potential are shown in fig.(2a)
and (2b).
The ground state
$f_{+}$ and the first excited state $f_{-}$ are shown in fig.(2c) and (2d).
They are represented
by spinors $(^{1}_{1})$ and $(^{1}_{-1})$ respectively, which
have spin vectors along $+{\bf \hat x}_{1}$
 and $-{\bf \hat x}_{1}$. Fig.(2e) and (2f) represent
the sum and difference of these states, which are localized on the right
and left well resp. These states are denoted as $f_{\uparrow}$ and
$f_{\downarrow}$, and correspond to the spinors $(^{1}_{0})$ and
$(^{0}_{1})$. The corresponding spin vectors are ${\bf \hat{x}}_{3}$ and
$-{\bf \hat{x}}_{3}$.

\vspace{0.2in}

\noindent Figure 3:
The pseudo-spin representation of the crystals shown in figure 1.
The one-component hexagonal crystal in Figure 1c is represented by the
ferromagnet hexagonal crystal in figure 3a, with magnetization along
${\bf \hat{x}}_{1}$. The centered square structure in figure 1b is
represented by an antiferromagnet crystal in figure 3b, with sublattice
magnetization along  $+{\bf \hat x}_{3}$ and $-{\bf \hat{x}}_{3}$,
which are represented by solid and empty circles respectively.
This structure only exist at
zero tunneling $\Delta=0$. When $\Delta\neq 0$, both up and
down spins will tilt toward ${\bf \hat{x}}_{1}$ as shown in figure (3b').
An arrow with a solid or (empty) circle attached to the base denotes a spin
with positive (negative) ${\bf \hat{x}}_{3}$ component.
The staggered hexagonal
structure in figure 1a is represented by the antiferromagnetic structure in
figure (3c). Like  (3b), (3c) only exists at zero tunneling.
Nonzero tunneling will make the spins point along ${\bf \hat{x}}_{1}$ as
shown in (3c'). For later use, the structures in (3a), (3b'), and (3c') are
denoted as {\bf I}, {\bf II}, and {\bf III} respectively. (See figure 4a to
4c).

\vspace{0.2in}

\noindent Figure 4 :
Figure (4a), (4b), and (4c) show the
Wigner crystal phase diagram in the the plane of
$\Delta/(e^{2}/a)$ and $D/a$ at
filling factors $\nu=1/3, 1/5$ and $0$, where $a^{2}$ is the area of the
unit cell in lattice A.
Although the system is a quantum Hall fluid at these fillings, the Wigner
crystal phase diagram remains essentially the same at nearby fillings,
where the system is no longer a quantum Hall fluid.
The solid and dashed lines are first and second
order lines respectively.
The label (F) means ferromagnetic order, (M) means mixed ferromagnetic and
antiferromagnetic order.
Both spin and lattice structure undergo
discontinuous change across the first order line.
The lattice structure in {\bf I}, {\bf III},
and {\bf V} are ``rigid", in the sense that they are unchanged in the
entire {\bf I}, {\bf III}, and {\bf V} region.
In region {\bf II} and {\bf IV}, the lattice structure (i.e.
${\bf a}_{1}, {\bf a}_{2}$, and ${\bf c}$) varies with
$\Delta/(e^{2}/a)$ and $D/a$.
The spin structure is only ``rigid" in region {\bf I}
where it points along ${\bf \hat{x}}_{1}$. In all other regions, the spin
angle varies with $\Delta/(e^{2}/a)$ and $D/a$.
See Section III for the
determination of  the spin angle $\theta$.

\vspace{0.2in}

\noindent Figure 5 : Figure 5a to 5d show the Wigner crystal phase diagram
as a function of $\overline{\Delta}\equiv\Delta/(e^{2}/\ell)$ and $\nu$
for $D/\ell= 3, 2, 1, 1/2$. When plotted in these variables, the
phase boundaries show more  changes when compared with those
in figure 4a to 4c. In figure 5a, both phase {\bf II} and the multicritical
point are not shown for they
appear at very small $\overline{\Delta}$ and $\nu$. They are shown in
figure 5b, 5c, and 5d. The spin angle $\theta$ along the
vertical and the horizontal line are shown in Figure 7 and 8.

\vspace{0.2in}

\noindent Figure 6 : Figure 6a and 6b show the centered rectangular
and centered rhombic structure, (i.e. crystal {\bf II} and {\bf IV}) in
figure 4a to 4c. The meaning of the arrows is identical to that in the
caption of figure 3. The crystal (fig.6a) reduces to that in fig. 3a
when $a_{2}/a_{1}=\sqrt{3}$, and $\theta=90^{o}$.

\vspace{0.2in}

\noindent Figure 7 : The spin angle $\theta$ along the vertical line in
figure 5a. There is a close to $20^{o}$ discontinuity across the
{\bf I}-{\bf III} first order line, where as a very small discontinuity
(almost invisible on this scale) across the {\bf IV}-{\bf V} first order
line.

\vspace{0.2in}

\noindent Figure 8 : The spin angle $\theta$ along the horizontal line in
figure 5a. The discontinuities of the spin angle across the
{\bf I}-{\bf III} and {\bf IV}-{\bf V} first order lines are similar to
that in figure 7.

\vspace{0.2in}

\noindent Figure 9 : Figure 9a to 9d show  $\gamma^{\bf I},
\gamma^{\bf III}, \gamma^{\bf V}$ as a function of $\nu$ for
$D/\ell$=3, 2, 1, 1/2. Using eq.(\ref{ang}), these curves allow one to
determine the
spin angle $\theta$ for the crystals  {\bf I}, {\bf III}, and {\bf V} in
figure 5a to 5d.

\vspace{0.2in}

\noindent Figure 10 : Figure 10(a) shows the energies of the crystal
states ${\bf I}$ to
${\bf V}$ at $\nu=1/3$ and $\Delta/(e^{2}/a)=0.1$ as a function of
layer separation $D/a$. Figure 10(b) shows the transition region
${\bf I}$-{\bf II}
and {\bf II}-{\bf III} in figure 10(a) on an expanded scale. The
energy difference depicted is typically of the order of $10^{-3}$.


\begin{references}
\bibitem{ATT1} Y.W. Suen, L.W. Engel, M.B. Santos, M. Shayegan, D.C. Tsui,
 Phys. Rev. Lett. {\bf 68}, 1379 (1992); J.P. Eisenstein, G.S.
Boebinger, L.N. Pfeiffer, K.W. West, and Song He,
Phys. Rev. Lett. {\bf 68}, 1383
(1992).  S.Q. Murphy, J.P. Einstein, G.S. Boebinger, L.N.
Pfeiffer, and K.W. West, Phys. Rev. Lett. {\bf 72}, 728 (1994).
Y.W. Suen, H.C. Manoharan, X. Ying,
M.B. Santos, M. Shayegan,
 Phys. Rev. Lett. {\bf 72}, 3405 (1994).

\bibitem{QHF} K. Yang, K. Moon, L. Zhang, A.H. MacDonald, S.M.
Girvin, and S.C. Zhang, Phys. Rev. Lett. {\bf 72}, 732 (1994).

\bibitem{Ho} T.L. Ho, Phys. Rev. Lett. {\bf 73}, 874 (1994).

\bibitem{wc} V.J. Goldman, M. Santos, M. Shayegan, and J.E. Cunningham,
Phys. Rev. Lett. {\bf 65}, 2189 (1990).

\bibitem{wcbell} H.W. Jiang, R.L. Willett, H. L. Stormer, D.C. Tsui, L.N.
Pfeiff
er,
K.W. West, Phys. Rev. Lett. {\bf 65} 663 (1990).

\bibitem{thre}  Y.P. Li, T. Sajoto, L.W. Engel, D.C. Tsui, and M. Shayegan,
Phys. Rev. Lett. {\bf 67}, 1630, (1991).
F.I. Williams, P.A. Wright, R.G. Clark, E. Y. Andrei,
G. Deville, D.C. Glattli, O. Probst, B. Etienne,
C. Dorin, C.T. Foxon, and J.J. Harris,  Phys. Rev. Lett. {\bf 66}, 3285 (1991).

\bibitem{magmode}
E. Y. Andrei, G. Deville, D.C. Glattli, F.I. B. Williams, E. Paris, and
B. Etienne, Phys. Rev. Lett. {\bf 60}, 2765 (1988); F. I. B. Williams,
{\em et al } ibid {\bf 66}, 3285 (1991).

\bibitem{lum}
H. Buhmann, W. Joss, K. von Klitzing, I.V. Kukushkin, G. Martinez,
A.S. Plaut, K. Ploog, and V. B. Timofeev, Phys. Rev. Lett. {\bf 65}, 1056
(1990).

\bibitem{ins} S. Kivelson, D.H. Lee, S.C. Zhang, Phys. Rev. {\bf B 46},
2223, (1992).


\bibitem{MZ} K. Maki and Zotos, Phys. Rev. {\bf B 28}, 4349 (1983).

\bibitem{BM} L. Bonshall and A. Maradudin, Phy. Rev. B, {\bf 15}, 1959 (1977).

\bibitem{num} See for example, W.H. Press, B.P. Flannery, S.A. Teukolsky, and
W.T. Vetterling, {\em Numerical Recipes}, Cambridge University Press, 1986.

\bibitem{ZF} L. Zheng and H.A. Fertig, to be published.
\end{references}
\end{document}